\documentclass[11pt]{article}
\usepackage{fullpage}
\usepackage{times}
\usepackage{latexsym}
\usepackage{amsmath}
\usepackage{color}
\usepackage{graphicx}
\usepackage{subcaption}
\usepackage{xspace}
\usepackage[authoryear,round,longnamesfirst]{natbib} %
\usepackage{booktabs}
\usepackage{multirow} %
\usepackage{enumitem}
\usepackage[normalem]{ulem}
\usepackage{hyperref}

\renewcommand{\cite}[1]{\citep{#1}}

\newcommand{\para}[1]{\smallskip\noindent{\bf #1}}
\newcommand{\addFigure}[2]{\includegraphics[width=#1]{plots/#2}}

\renewcommand{\cite}{\citep} %

\newif{\ifhidecomments}
\hidecommentsfalse   %

\ifhidecomments
   \newcommand{\llee}[1]{}
   \newcommand{\chenhao}[1]{}
\else
  \newcommand{\llee}[1]{\textcolor{red}{#1}}
  \newcommand{\chenhao}[1]{\textcolor{blue}{#1}}
\fi

\newcommand{\context}[1]{context\xspace}
\newcommand{\contextword}[1]{\textbf{\textcolor{red}{\underline{#1}}}}
\newcommand{\inword}[1]{\textit{\textcolor{blue}{#1}}}
\newcommand{\outword}[1]{\textit{\textcolor{cyan}{#1}}}
\newcommand{\wordpair}[2]{(\inword{#1}, \outword{#2})}
\newcommand{\hedge}{\contextword}
\newcommand{\future}{subsequent\xspace}

\newcommand{\negconj}{{\em contrastive conjunctions}\xspace}
\newcommand{\Negconj}{{\em Contrastive conjunctions}\xspace}

\newcommand{\secondperson}{{\em second person pronouns}\xspace}
\newcommand{\Secondperson}{{\em Second person pronouns}\xspace}

\newcommand{\shielding}{{\em hedges}\xspace}
\newcommand{\Shielding}{{\em Hedges}\xspace}
\newcommand{\shieldingnoemph}{hedges\xspace} %
\newcommand{\superlative}{{\em superlatives}\xspace}
\newcommand{\Superlative}{{\em Superlatives}\xspace}

\newcommand{\secref}[1]{\S \ref{#1}}
\newcommand{\tableref}[1]{Table \ref{#1}}
\newcommand{\figref}[1]{Fig. \ref{#1}}

\newcommand{\speaker}[1]{{\em #1}} %

\title{
{ %
Talk it up or play it down? \\
(Un)expected correlations
between (de-)emphasis
and
recurrence
of discussion points
in
consequential
U.S. economic policy
meetings
}}

\author{Chenhao Tan\footnote{Work done while the author was at Cornell University.}\\
  Dept. of Computer Science \\
  University of Washington \\
  {\url{https://chenhaot.com}}
  \medskip\\
  Lillian Lee \\
  Dept. of Computer Science \\
  Cornell University \\
  {\url{http://www.cs.cornell.edu/home/llee}}\\
}

\date{
Presented at Text as Data, Oct 14-15, 2016 \\
Comments welcome; please send to chenhao@chenhaot.com and llee@cs.cornell.edu
}

\begin{document}
\maketitle

\begin{abstract}
In meetings where important decisions get made, what items receive
more attention may influence the outcome.
We examine how different types
of
rhetorical \mbox{(de-)emphasis}  ---
including hedges,
superlatives,
and
contrastive conjunctions
---
 correlate with what
gets revisited
later,
controlling for item frequency and speaker. Our data consists of transcripts
of recurring
meetings
of the Federal Reserve's Open Market Committee (FOMC),
where
important
aspects of U.S. monetary policy are decided on.
Surprisingly, we find that words appearing in the context of hedging, which is usually considered a way to express uncertainty, are more likely to be repeated in subsequent meetings, while strong emphasis indicated by \superlative has a slightly negative effect on word recurrence in \future meetings.
We also observe interesting patterns in how these effects vary depending on social factors such as status and gender of the speaker.
For instance, the positive effects of hedging are more pronounced for female speakers than for male speakers.

\end{abstract}

\section{Introduction}
\label{sec:intro}

Meetings play a crucial role in a wide range of settings, including collaboration, negotiation and
policy decisions \cite{Jarzabkowski01112008}.
For example, the U.S. Federal Open Market Committee (FOMC),
``the monetary policymaking body of the Federal Reserve System'',\footnote{
{\url{https://www.federalreserve.gov/faqs/about_12844.htm}}}
``holds eight regularly scheduled
[six-hour]
meetings per year [where it] reviews economic and financial conditions, determines the appropriate
stance of monetary policy, and assesses the risks to its long-run goals of price stability and sustainable economic growth''; its decisions can ``ultimately [affect] a range of economic variables, including employment, output, and prices of goods and services''.\footnote{{
\url{https://www.federalreserve.gov/monetarypolicy/fomc.htm}}}
Studies of the FOMC's meetings or using FOMC meeting transcripts as
data include \citet{meade2005fomc,Meade:TheEconomicJournal:2008,
SchonhardtBailey:DeliberatingAmericanMonetaryPolicyATextualAnalysis:2013,guo2015bayesian,zirnlost,hansentransparency}.

\subsection{Hedging ``versus'' superlatives}
A central question for each meeting participant is how to make
his or
her
arguments
noted and valued by other participants, and
thus ultimately
influence the outcome of the meeting.
Our interest in this paper is in the effectiveness of certain
subtle
presentational or
rhetorical options in this regard --- specifically, whether a speaker
attempts to make a point using certain vs.~uncertain language.
Here is an
example taken from the March 22, 2005 FOMC meeting. The speaker is
identified in
the transcript as Ms. Minehan,\footnote{The presence of ``Ms.'' and ``Mr.''
notations in the transcripts mean that we can easily extract gender
information, a fact we take advantage of in our experiments.} President of
the
Federal Reserve Bank of
Boston, and she is discussing an alternative wording:%
\footnote{We
acknowledge the meta-ness of including as an example in a paper about
choices of wording a case where people are discussing choices of wording.}
\begin{quote}
I'm also concerned in alternative B about the rise in energy prices not
notably feeding through to core consumer prices. Core consumer prices are up
a full percentage point on a year-over-year basis, and there has been some
feed-through. We think it's going to slacken, and {\em \hedge{maybe}
you want to put
that
reference in the future, but \hedge{I'm not sure} that this is what we want
to say
in this statement.} I think we'd be better off leaving that sentence out and
just going with ``pressures on inflation have picked up in recent months and
pricing power is more evident.''
\end{quote}
The italicized sentence contains the highlighted hedges ``maybe'' and
``I'm not sure''. Notice that Minehan could have uttered a more
invested or committed version of this
sentence
that omits the
expressions of uncertainty:
\begin{equation}
\ldots\mbox{ and you could  put that reference
in the
future, but this is not what we want}\ldots
\end{equation}
Also, she could have made the point using superlative language for emphasis:
\begin{equation}
\ldots
\mbox{this is the \hedge{worst} wording we could possibly pick.}
\end{equation}
Would one of these choices have been more effective than the others in
causing the committee members to seriously consider Minehan's proposals?

\bigskip

\subsection{Why hedging?}It may at first seem strange to choose the
``emphasis'' aspect of wording as a focal
point. One objection runs as follows: besides wording, there
are many other, perhaps more salient
factors at play, such as status, social relationships, shared history,
charisma,
timing,
and so on \cite{Cialdini:InfluenceScienceAndPractice:2009}, not to mention the validity or ``correctness'' of the
content of
an argument itself \cite
{Petty:CommunicationAndPersuasionCentralAndPeripheralRoutes:2012}.
However, the ``omnipresence''
of
the idea of {\em framing} ``across the social sciences and
humanities''
 means that there is a great deal
 of scholarly interest in how speakers and authors can, {\em often through
 language}, ``select some aspects of a perceived reality and make them more
 salient'' for persuasive ends \cite{Entman:JournalOfCommunication:1993}.
Moreover, we have argued elsewhere that
{\em how someone \underline{says} something is one of the few factors that
a speaker has some control over
when
he or she seeks to convey a fixed piece of content}:
\begin{quote}
For example, consider a speaker at the ACL [a scientific organization's]
business meeting who has been
tasked with proposing that Paris be the next ACL location. This person
cannot on the spot become ACL president, change the shape of his/her social
network, wait until the next morning to speak, or campaign for Rome instead;
but he/she can craft [their] message ...
\citep{Tan:Acl:2014}
\end{quote}
We thus assert that it is both an interesting scientific question and an
interesting pragmatic question to ascertain
whether
language aspects of delivery have an effect on the degree of influence one
has, independent of non-linguistic factors.

Our particular interest in this paper in looking at employment of
expressions of
uncertainty arises from how fascinating the phenomenon is in its own right
(see, for example,
\citet{Schroeder:HedgingInDiscourseApproachesToTheAnalysis:1997} for a
listing of perhaps hundreds of papers on the topic up to 1997, and \citet
{Farkas+al:2010a} for a description of entrants to a  shared
task/competition among NLP systems for identifying uncertainty).  After all,
the fact that hedging exists is seemingly odd:
one
might first think that if
people want communication to be direct and efficient, shouldn't they just
cut out the extra verbiage that hedging entails? And, don't hedges make a
speaker
or a speaker's position
seem weak? Public-speaking advice on the Internet cautions
people to avoid them, and indeed,  Strunk and White themselves state:
``Avoid tame, colorless, hesitant,
non-committal language.''

But in fact hedging can be a
tool for a speaker to achieve his or her aims.  Consider the following
excerpt from the March 22, 2005 FOMC meeting, where Kos hedges much more
than Greenspan, the chair:
\begin{equation}
\begin{minipage}[t]{6in}
\speaker{Greenspan}: \contextword{I assume} iron ore is in [the CRB]? \\
  \speaker{Kos}: {I don't know} if iron ore is in there
  but copper is: copper scrap is in there, \contextword{I think}. \\
\speaker{Greenspan}: That couldn't have done that much. Steel, for
  example, is actually down. \\
   \speaker{Kos}: \contextword{I don't think} steel is in the CRB.
\end{minipage}
\end{equation}
%
%
%
%
%
%
%
%
%
%
%
Importantly, Kos's corrections of Greenspan are accurate:
according to Thomson Reuters\footnote{
\url{http://financial.thomsonreuters.com/content/dam/openweb/documents/pdf/financial/core-commodity-crb-index.pdf}}, the CRB index
contains copper but not
iron ore or steel. Furthermore, Kos is presumably not actually uncertain of
these facts.  Rather, it would seem that Kos is softening his language
to either (1)
make his assertions more palatable or acceptable, or (2)
trying to signal respect while contradicting the higher-status Chair.

\subsubsection{Why the FOMC?}

So far, we have not mentioned anything about language usage that seems
particularly specific to economic policy discussion.
But the FOMC meetings are a particularly
nice domain for our empirical work because we
might expect language effects to be minimized:
\begin{itemize}
\item The stakes are very high, since the decisions made by the FOMC are
extremely consequential. Thus, one might argue that the
participants would
be highly motivated to focus on the content, not the wording, of the
discussions.
\item The participants are high-status experts in the field, and
hopefully respect each other to at least some extent.  One might therefore
suppose that they would
be less inclined to either require expressions of social deference to each
other or be impressed by undeserved emphaticism,
especially as the meetings wear on over multiple hours.
\item At least some of the participants have interacted a great deal with
each other, which might reduce the influence that language choices would
have on how people's suggestions are received by each other.
\end{itemize}
Hence, since the situation reduces the possibility for
language choices to have an effect, any effects that we {\em do} see deserve
consideration.

Moreover, some other experimentally convenient features are (a) the
positions (job descriptions) and genders of the
participants are indicated in the transcripts; and (b) pre-1993, the FOMC
members were not aware that the transcripts would
be made public --- this fact dampens the possibility that the participants
were speaking unnaturally or trying to direct their comments towards a
broader audience.
Other characteristics we
have not exploited in this work but could be useful for other research
include (c) many speakers participate in many meetings, providing
relatively plentiful user-specific data; (d) there is a great deal of
public documentation laying out the basis on which decisions are made and
what is being decided upon, such
as the
Bluebooks, Greenbooks, and so on;\footnote{\url
{https://www.federalreserve.gov/monetarypolicy/fomc_historical.htm}}
(e) \citet{meade2005fomc} provides manually-assigned disagreement labels
which indicate
who
argued against --- not just cast a dissenting vote against ---  the final
decision for each meeting,
which may be interesting for future studies.
We intend to make our processed versions of the
transcripts publicly available.

\subsection{A repetition framework for investigation }

While we would like to study whether
hedging
and other forms of (de-)emphasis
have detrimental or positive
effect on the reception of a speaker's ideas, it can be difficult to
ascertain (computationally or otherwise) whether the listeners give those
ideas serious consideration or not.

We therefore employ the following computationally convenient {proxy} for
idea uptake: {\em repetition} or {\em echoing}, inspired by \citet{Niederhoffer+Pennebaker:2002a,DanescuNiculescuMizil:ProceedingsOfWww:2011a,Danescu-Niculescu-Mizil:2012:EPL:2187836.2187931}. (See also the definition of ``discussion points'' of \citet
{Zhang:ProceedingsOfNaacl:2016}.)

\figref{tb:example} demonstrates the main idea
of how we construct the specific data-points for our study.
For a given context, such as ``expressions of uncertainty'' or
``superlatives'', we find instances of the occurrence of the context in
individual speeches.\footnote{By ``speech'', we mean an uninterrupted span
of
speech by a single speaker.} Then, we pair a word from the speech appearing
outside
the
context with a word from the speech appearing in
the
context word or phrase, taking care that the ``in''-word has the same
frequency prior to the speech as the ``out''-word.

We then ask, how
frequently does the out-of-context word occur
{\em after the speech}, in comparison to the in-context word --- we thus use
``a word is used by other people'' as a rough proxy for ``other people are
paying attention to the underlying concept''.
The
null hypothesis, given that both words are of equal prior
probability and uttered by the same speaker at almost the same time, is that
context will have no effect and the words will continue to have roughly
equal frequency in the future.
By definition this framework controls for important factors other than the phrasing,
such as who the speaker is and
when in the meeting does
the speech happen.

Note that
our framework also allows for
flexibility in
measuring how well  statements are received:
using repetition as the
indicator of influence is not central to
our setup.

\subsection{Highlights reel}

Our first contribution is the repetition-based speaker- and time-controlled
framework we introduced in the previous subsection.

We look for hedging/emphasis-mediated repetition effects in the FOMC meeting
transcripts both within the same meeting (intra-meeting)
and in \future meetings
(inter-meeting).
One surprising finding is that although \shieldingnoemph have
very little
effect within the same meeting, words in the context of \shieldingnoemph are
more likely to occur in \future meetings.

\begin{figure}[t]
\centering
\fbox{\begin{minipage}[t]{6in}
\speaker{Mr. Moskow}: ...
Auto and light truck sales appear to
be coming in at about the 14-1/2 million units level so far in May,
which is approximately 3/4 million units above the April pace but
still well below the \outword{expectations} earlier this year.
[...]
On the employment front, labor markets remain tight, with the
District's \inword{unemployment} rate at its \contextword{lowest} level in over 15 years. ...
\end{minipage}
}
\caption{Example of a matching word pair \wordpair{unemployment}{expectations}
for the context of
\superlative (``\contextword{lowest}'' highlighted
and underlined in red).
They have been uttered by the same speaker, and are of similar frequency in
our dataset before this speech;  and ``unemployment'' occurs in the context of ``lowest'', while ``expectations'' does not.
}
\label{tb:example}
\end{figure}

Furthermore, we investigate how these effects may vary depending on other factors such as status and gender of the speaker.
One interesting finding is that the effect of context is more pronounced
for female speakers.
This echoes existing work that suggests that female speakers are more persuasive in an indirect manner \cite{burgoon1975toward}.
\section{Analysis framework}
\label{sec:exp}

\newcommand{\incontext}{in}
\newcommand{\outcontext}{out}

In order to understand the effect of
(de-)emphasis
on the reception --- where here we approximate ``reception'' by ``repetition'' ---  of a speaker's ideas,
we develop a framework that
controls for important confounding factors.
Throughout, we use the term {\em context} to refer generically to a class of
(de-)emphasis techniques.

The intuition is shown in \figref{tb:example}: for a given context, within the same speech
--- so that
both the speaker and the
meeting state are naturally controlled ---
we identify sentences containing an instance of that context
versus sentences that do not contain any instances of that context.
We then extract ``similar'' (in-context, out-of-context) word pairs
based on their frequency in the past, and compare their
frequency of occurrence later in the meeting or in subsequent meetings.

\para{Formal definition.}
A context $C$ is defined as a set of words or phrases.
Within the same speech $S$,
we define sentences that contain
any
$c \in C$ as $Sent_{\incontext}$, and the other sentences as $Sent_{\outcontext}$.
We match content words of similar past frequency in $Sent_{\incontext}$ and $Sent_{\outcontext}$,\footnote{We exclude stopwords and words in
all matching pairs,
although the results are robust even if we include stopwords in the pairing process.}
and define the set of matched pairs for speech $S$ as
\newcommand{\MP}{{\rm MP}_C\xspace}
 $\MP(S)$:
\begin{equation}
\MP(S) \stackrel{def}{=} %
\{(w_{\incontext}, w_{\outcontext}, S) \,\vert \, w_{\incontext} \in
{\scriptstyle Sent_{\incontext} - Sent_{\outcontext}},
w_{\outcontext} \in {\scriptstyle Sent_{\outcontext} - Sent_{\incontext}},
PF_S(w_{\incontext}) \sim PF_S(w_{\outcontext})\},
\label{eq:mp}
\end{equation}

\noindent where $PF_S$ gives the past frequency of a word.
Some further details on the finer points of refining the definition of $\MP$ are given in the appendix, \secref{sec:validation}.

Next, we define the {\em effect} of a context by measuring the difference between the probability that words in the context are echoed {\em more} in the future than out-of-context words and a default prediction of 0.5, since the words in our same-speech pairs have similar past frequency:
\begin{equation}
E(C) = \widehat{{P}}_C
- 0.5,
\label{eq:effect}
\end{equation}

\noindent where $\widehat{{P}}_C$ computes the probability that words in the context are repeated more in the future than words out of the context.
Specifically, we compute the average winning rate of $w_{\incontext}$ in $MP_C(S)$ for each speech $S$ and then average over all speeches:
$\widehat{{P}}_C=\frac{1}{|\mathcal{S}|}\sum_{S \in \mathcal{S}}\frac{1}{|MP_C(S)|}\sum_{w_{\incontext}, w_{\outcontext} \in MP_C(S)} I(FF_S(w_{\incontext}) > FF_S(w_{\outcontext}))$.
Here $FF_S$ gives the frequency of a word in {\em other meeting participants' speeches} after $S$ either within the same meeting or in \future meetings.
The definition of $FF_S$
can vary depending on research hypotheses that we are interested in.
We will present two classes of $FF_S$ in \secref{sec:data}.

{\em In our experiments, we will be concerned with whether the effect defined by Equation \eqref{eq:effect} is different from zero.}
A positive effect suggests that the context is associated with more future echoing, while a negative effect suggests less.
\section{Hypotheses}
Our main interest in this work is to examine the effect of \mbox{(de-)emphasis} on the reception of a speaker's ideas.
In addition to \shielding and \superlative, we also investigate two other common contexts that can be associated with emphasis: \negconj and \secondperson.
In the following, we develop our hypotheses based on existing studies and our intuitions.

\begin{itemize}

\item[H1:] \Shielding.
We have already discussed some intuitions and prior work regarding hedging in the Introduction.  Moreover,
\citet{Durik:JournalOfLanguageAndSocialPsychology:2008} show that ``hedges can, but do not always, undermine persuasive attempts'' and \citet{Erickson+al:78a} show that powerless language results in lower perceived credibility of the witness in court trials.
In light of these studies, we expect a {\em negative} effect within a meeting.
We merge and manually curate several data sources to get a list of hedges \cite{Farkas+al:2010a,hanauer2012hedging,hyland1998hedging}.\footnote{We focus on a subset of hedges where the speaker may try to shield the responsibility of a statement. For example, ``to be raised'' and ``or'' from \citet{Farkas+al:2010a} are not included.}
\footnote{It should be pointed out that the automatic identification of hedging and expressions of uncertainty is not a solved problem \cite{Farkas+al:2010a}. Items that seem like hedge cues may turn out always be so in real-life usage (compare ``I {\bf *think*} it's going to rain'' with ``{\bf *I*} think it's going to rain'); and, hedging can occur without well-recognized hedge cues (``I'm no Albert Einstein, but I say the answer is 1234.'').}

\item[H2:] \Superlative.
As these are the strongest form to describe a fact or an action and can place an emphasis on the statement\footnote{This sentence itself contains a superlative: the word ``strongest''.},
we expect a {\em positive} effect.

\item[H3:] \Negconj.
A contrastive conjunction like ``but'' places
an emphasis on the text after its occurrence,
so we expect a {\em positive} effect.
\item[H4:] \Secondperson.
Although using \secondperson (``you'') is not a form of emphasis, it can likely attract the attention of the addressed speaker.
We expect a {\em positive} effect shortly after the speech as these are direct mentions of other meeting participants.
\item[H5:] No lasting effects.  We expect that so much time passes between
meetings and the word choices we are looking at are sufficiently subtle
(for instance, the addition of the phrase ``I think') 'that there should be
no effects lasting from one meeting to the next.

\end{itemize}

\section{Dataset}
\label{sec:data}

\begin{table}[t]
\centering
\small
\begin{tabular}{rrr}
\toprule
  Num. speakers & Num. speeches &  Num. tokens \\
\midrule
26.93 & 404.87 & 37652.40 \\
\bottomrule
\end{tabular}

\normalsize
\caption{Per-meeting averages in our dataset. \label{tb:basic_stat}}
\end{table}

Our dataset is drawn from the transcripts of all FOMC meetings from 1977 to 2008.
\tableref{tb:basic_stat} presents basic statistics.
In order to apply our framework, we define the {\em past frequency} of a word ($PF_S$)
with respect to its appearance in a speech $S$
as the log probability of the word in the previous meetings,
and employ two classes of functions to measure the {\em future frequency} of a word:

\begin{itemize}
\item Intra-meeting frequency.
We split the speeches after $S$ into windows of five speeches,
and then compute the log probability of a word within each window after $S$ for 20 windows (100 speeches after $S$).
We expect the effect
of a context
to fade away as the meeting moves forward, whether
that effect is
positive or negative.

\item Inter-meeting frequency.
In order to assess the effect of (de-)emphasis in \future meetings, we compute
per-meeting log probability of the word
for each of the five \future meetings after $S$.
\end{itemize}

\begin{table}[t]
\centering
\small
\begin{tabular}{lp{0.5\textwidth}}
\toprule
context & example pairs \\
\midrule
\shielding & \wordpair{adjustment}{capacity}, \wordpair{mortgage-related}{high-yielding} \\
\midrule
\superlative & \wordpair{sector}{employment}, \wordpair{information}{judgment}\\
\midrule
\negconj & \wordpair{money}{point}, \wordpair{export}{signal}\\
\midrule
\secondperson & \wordpair{disturbance}{shipments}, \wordpair{spreads}{reductions}\\
\bottomrule
\end{tabular}

\normalsize
\caption{Example pairs for different contexts.
The first element in each pair is the in-context word; the second is the
outside-context word.
Recall that these are words spoken by the {\em
same} speaker at {\em about the same time}. \label{tb:pair_example}}
\end{table}

Recall that we compare the future frequency of prior-frequency-controlled (in-context, outside-context) pairs.
\tableref{tb:pair_example} presents two matching pairs of words randomly chosen from our pairs data for each of the four contexts that form the foci of our hypotheses.
Indeed, it is non-trivial to guess which word, if any,  will be echoed significantly more
in the future
a priori.

As a preliminary experiment,
we used the method of \citet{Monroe21092008} to
compare
the words tending to appear within each type of context with the words tending to appear outside each type of context.
We omit detailed results here, but
in general,
the differences match our intuitions.
For example, \shielding tend to occur with evaluative statements (``ought'', ``risks'', ``important''),
while
``thank'' and ``chairman'' typically occur out of context,
because a typical phrase to start a speech in these meetings is ``Thank you, Mr. Chairman.''
But, recall that we purposely constructed pairs to have equal prior probability,
which should help mitigate any effects stemming merely from what words tend to occur in a given context.

\begin{figure*}[htb!]
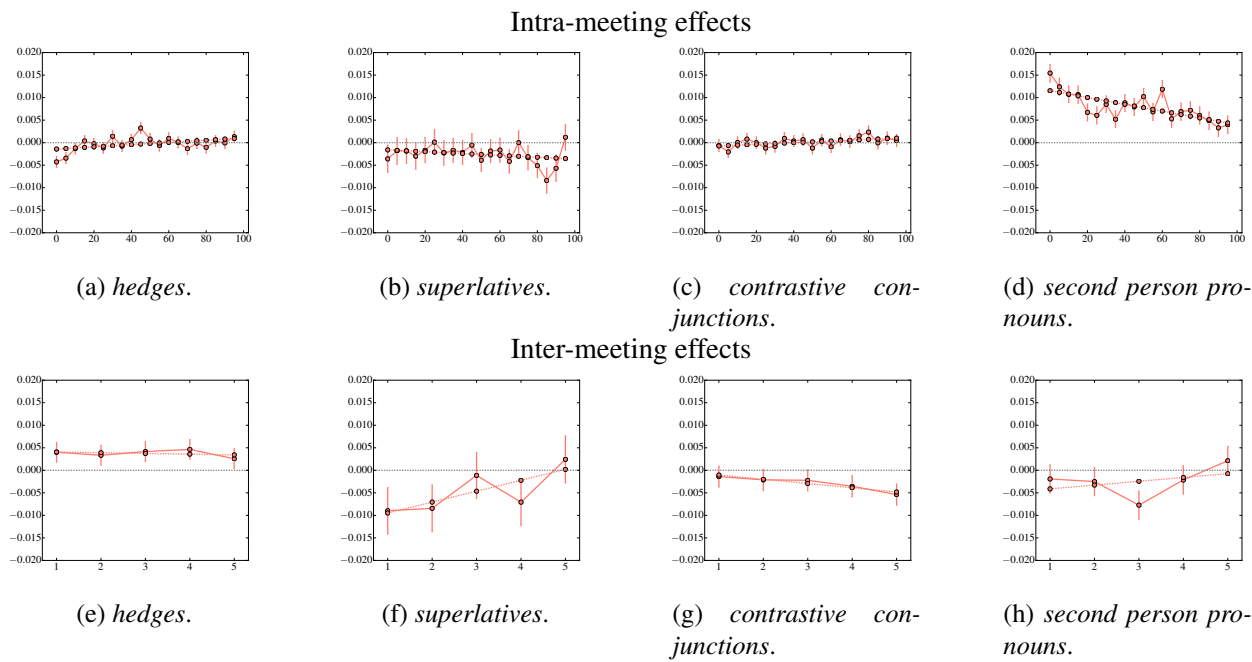

\centering
	Intra-meeting effects\\
	\begin{subfigure}[t]{0.2\textwidth}
		\addFigure{\textwidth}{{infreq_intra_shielding_0.010000_noself_1006_in_out}.pdf}
		\caption{\shielding.}
		\label{fig:intra_shielding}
	\end{subfigure}
	\hfill
	\begin{subfigure}[t]{0.2\textwidth}
        \addFigure{\textwidth}{{infreq_intra_super_0.010000_noself_1006_in_out}.pdf}
        \caption{\superlative.}
        \label{fig:intra_super}
    \end{subfigure}
    \hfill
    \begin{subfigure}[t]{0.2\textwidth}
		\addFigure{\textwidth}{{infreq_intra_negative_conj_0.010000_noself_1006_in_out}.pdf}
		\caption{\negconj.}
		\label{fig:intra_contrast}
	\end{subfigure}
	\hfill
    \begin{subfigure}[t]{0.2\textwidth}
		\addFigure{\textwidth}{{infreq_intra_you_0.010000_noself_1006_in_out}.pdf}
		\caption{\secondperson.}
		\label{fig:intra_you}
	\end{subfigure}\\
    Inter-meeting effects\\
    \begin{subfigure}[t]{0.2\textwidth}
		\addFigure{\textwidth}{{infreq_inter_shielding_0.010000_noself_1006_in_out}.pdf}
		\caption{\shielding.}
		\label{fig:inter_shielding}
	\end{subfigure}
	\hfill
	\begin{subfigure}[t]{0.2\textwidth}
        \addFigure{\textwidth}{{infreq_inter_super_0.010000_noself_1006_in_out}.pdf}
        \caption{\superlative.}
        \label{fig:inter_super}
    \end{subfigure}
    \hfill
    \begin{subfigure}[t]{0.2\textwidth}
		\addFigure{\textwidth}{{infreq_inter_negative_conj_0.010000_noself_1006_in_out}.pdf}
		\caption{\negconj.}
		\label{fig:inter_contrast}
	\end{subfigure}
	\hfill
    \begin{subfigure}[t]{0.2\textwidth}
		\addFigure{\textwidth}{{infreq_inter_you_0.010000_noself_1006_in_out}.pdf}
		\caption{\secondperson.}
		\label{fig:inter_you}
	\end{subfigure}
	\caption{
	The first row presents the effect of contexts with the same meeting over 100 speeches (20 windows) after
	speech $S$;
	the second row presents the effects over five \future meetings.
	Dotted red line: the best linear fit of the effect over the x-axis.
	Dotted gray line: the null hypothesis where the context has no effect.
    x-axis: the number of speeches after the speech of interest in intra-meeting plots, the subsequent x-th meeting in inter-meeting plots.
    Bars: standard error.
    We use the same x-axis, y-axis, line styles, and bars in all intra-meeting and inter-meeting figures.
	\label{fig:overall}}
\end{figure*}

\section{Effects of contexts}
\label{sec:results}

We first examine the overall effects of contexts, and then explore how the effects vary across different factors,
including
status, gender, and speech length.

\subsection{Overall effects (\figref{fig:overall})}

\para{Intra-meeting effects (Hypotheses H1-H4).} We expected
\shielding (Hypothesis H1) to have a negative effect within the same meeting,
given that material that people express uncertainty about might tend to receive less attention
from the other participants.
However, \shielding seem to only have a small negative effect right after the speech and the effect quickly returns to 0.\footnote{%
Those of us who find ourselves tending to hedge may view this as an unexpectedly positive finding.
}

We also expected  that \secondperson, \negconj and \superlative (Hypotheses H2, H3, and H4)
would have a positive effect shortly after the speech of interest.
However, \superlative seem to have a slightly negative effect, while \negconj do not have much effect.
In contrast,
perhaps because
\secondperson directly mention other participants, they demonstrate a strong positive effect,
although, not surprisingly, the effect
diminishes over the course of the meeting.

\para{Inter-meeting effects (Hypothesis H5)}
In contrast with intra-meeting results, surprisingly,
\shielding correlate with a positive effect in the \future meetings.
This suggests that expressing uncertainty correlates with a better reception of ideas in the long run indicated by repetition.

Consistent with intra-meeting results, \superlative present a negative effect on whether words are going to be repeated in \future meetings.
Although \negconj present zero effect in the next several \future meetings, they lead to slightly more pronounced negative effect in later \future meetings.
Finally, the effect of \secondperson mostly overlaps with the zero line (in fact, it is very similar to the random case shown in \secref{sec:validation}).

\subsection{Impact of different factors}

Despite the above aggregate results,
the effect of a context may not be homogeneous conditioned on other factors, such as status (whether the speaker is chair or not), gender (whether the speaker is male or female),
and speech length (whether the speech is long or short).
We explore these variations in the inter-meeting effect of hedging and in the intra-meeting effect of \secondperson.
\begin{figure*}[t]
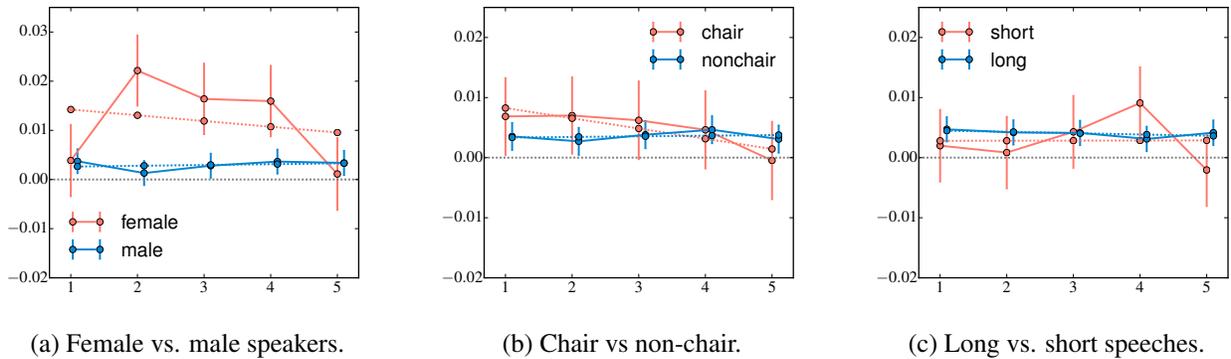

	\begin{subfigure}[t]{0.3\textwidth}
		\addFigure{\textwidth}{{nochair_gender_infreq_inter_shielding_0.010000_noself_1006_in_out}.pdf}
		\caption{Female vs. male speakers.}
		\label{fig:inter_gender_shielding}
	\end{subfigure}
	\hfill
	\begin{subfigure}[t]{0.3\textwidth}
		\addFigure{\textwidth}{{chairman_infreq_inter_shielding_0.010000_noself_1006_in_out}.pdf}
		\caption{Chair vs non-chair.}
		\label{fig:inter_chairman_shielding}
	\end{subfigure}
	\hfill
	\begin{subfigure}[t]{0.3\textwidth}
		\addFigure{\textwidth}{{length_infreq_inter_shielding_0.010000_noself_1006_in_out}.pdf}
		\caption{Long vs. short speeches.}
		\label{fig:inter_length_shielding}
	\end{subfigure}
	\caption{Inter-meeting comparisons of the
	effect of hedging across speaker status, gender and speech length. Note that the y-axis scale for gender comparison is different from the other two.\label{fig:impact-inter}}
\end{figure*}

\subsubsection{\Shielding (inter-meeting, \figref{fig:impact-inter})}
\label{sec:hedge_cmp}

\para{Stronger positive effect in \future meetings for female speakers. (\figref{fig:inter_gender_shielding})}
The gender of each participant can be obtained by the prefix in the speaker name.
We omitted all speeches made by the chairs to avoid the influence of status.\footnote{We tried to exclude Yellen and the same observation holds that there is stronger positive effect for female speakers.}
There is a clear positive effect in \future meetings for female speakers,
while there is not much effect for male speakers.
This echoes the findings in \citet{burgoon1975toward} and \citet{Carli:JournalOfSocialIssues:2001} that female speakers are more considered persuasive when employing an indirect manner.
 (For an interesting critique of advice that women should speak ``more like men'', see \citet{Cameron:VerbalHygieneInRecentYearsPractices:1995}.)

\para{Similar positive effect in \future meetings for speakers with different statuses. (\figref{fig:inter_chairman_shielding})}
We use whether the speaker was the chair of FOMC as a proxy of status.
As a result, the number of samples is much smaller for the chair group than the non-chair group and we thus observe a larger variance for the chairs.
The effect of \shielding for the chairs and non-chairs mostly overlap with each other, although the effect for the chairs seems to be slightly above that for non-chairs in the first several \future meetings.

\para{The positive effect in long speeches is more consistent. (\figref{fig:inter_length_shielding})}
The final aspect that we examine is speech length.
One may expect that for long speeches, it is more important to emphasize certain parts so that %
others can pick up.
To distinguish long speeches from short speeches, we simply split the speeches where there are matching word pairs into two groups using the median as a boundary.
The positive effect of \shielding in \future meetings is consistent for long speeches, while it fluctuates more for short speeches.

\begin{figure*}[t]
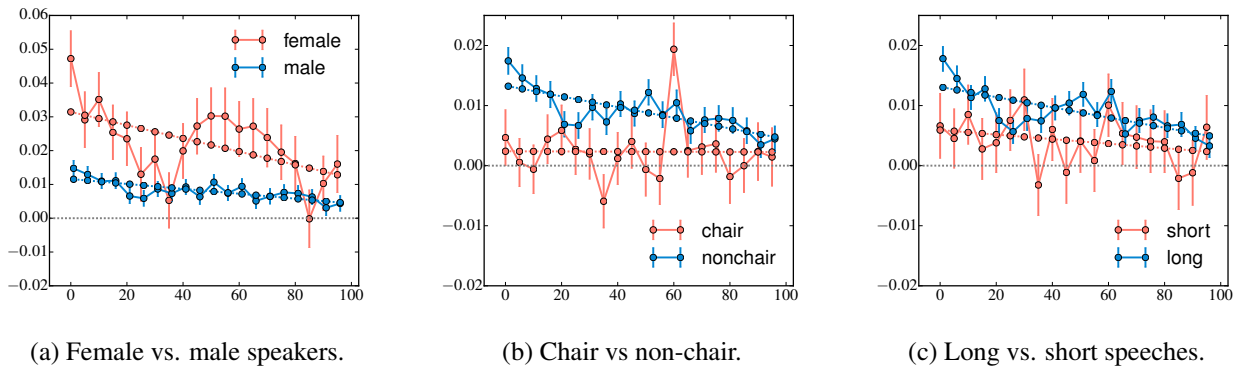

 	\centering
	\begin{subfigure}[t]{0.3\textwidth}
		\addFigure{\textwidth}{{nochair_gender_infreq_intra_you_0.010000_noself_1006_in_out}.pdf}
		\caption{Female vs. male speakers.}
		\label{fig:intra_gender_you}
	\end{subfigure}
	\hfill
	\begin{subfigure}[t]{0.3\textwidth}
		\addFigure{\textwidth}{{chairman_infreq_intra_you_0.010000_noself_1006_in_out}.pdf}
		\caption{Chair vs non-chair.}
		\label{fig:intra_chairman_you}
	\end{subfigure}
	\hfill
	\begin{subfigure}[t]{0.3\textwidth}
		\addFigure{\textwidth}{{length_infreq_intra_you_0.010000_noself_1006_in_out}.pdf}
		\caption{Long vs. short speeches.}
		\label{fig:intra_length_you}
	\end{subfigure}\\
\caption{Intra-meeting comparisons of the
effect of second-person pronouns across speaker status, gender and speech length.  Note that the y-axis scale for gender comparison is different from the other two.\label{fig:impact}}
\end{figure*}

\subsubsection{\Secondperson (intra-meeting, \figref{fig:impact})}
\label{sec:impact_you}

We examine how the positive effect of \secondperson within a meeting changes conditioned on status, gender and speech length.
We follow the same procedures as above to extract status, gender and speech length information.

\para{Stronger positive effect for female speakers than male speakers. (\figref{fig:intra_gender_you})}
Surprisingly, the effect of \secondperson is smaller for male speakers, in other words, \secondperson spoken by female speakers present a stronger immediate effect on other participants after the speech.
This may suggest that it is more important for female speakers to ``ask'' other participants to pay attention using certain contexts.
This observation is consistent with the result in inter-meeting results for \shielding:
the positive effect of a context is more pronounced for female speakers.

\para{Stronger positive effect for speakers with lower status than speakers with higher status. (\figref{fig:intra_chairman_you})}
The effect of \secondperson is mitigated in the chairs' speeches (y-values fluctuate around 0).
One way to interpret this observation is that meeting participants pay similar levels of attention to the chairs' statements regardless of the
second-person-pronoun
context.

\para{Stronger positive effect in long speeches. (\figref{fig:intra_length_you})}
The difference between long speeches and short speeches is clearer than in \secref{sec:hedge_cmp}.
The effect of \secondperson is stronger in long speeches.

\subsection{Further caveats and disclaimers}
We do not claim that correlation implies causation.  In particular, these findings should not be viewed as positive advice on how to influence discussion.

There are some aspects of the data that we do not  directly take into account in
the experiments reported in this paper.
There are changes over time in the style and leadership of the meetings.
For instance, the number of speeches is decreasing over time.
Also, after 1993, the meeting participants were aware of the fact that the
transcripts would be be made publicly available.
{}

\section{Related work}
\label{sec:related}

FOMC meetings have attracted significant research interests.
\citet{rosa2013financial} shows that the release of FOMC minutes significantly affects the volatility of U.S. asset prices and trading volume.
See \citet{SchonhardtBailey:DeliberatingAmericanMonetaryPolicyATextualAnalysis:2013} for a comprehensive textual analysis.
The focus of our study is on the effect of subtle rhetorical correlates within the meetings.

Another related line of work is accommodation and linguistic style matching \cite{Danescu-Niculescu-Mizil:2011:CIC:2021096.2021105,Niederhoffer+Pennebaker:2002a}, which study the phenomenon of people matching each other in conversations.
Here we attempt to study how
subtle
presentational and rhetorical (de-)emphasis
may influence the reception of a speaker's ideas and evaluate based on content words, in contrast with functional words to capture style.

Additionally, there have been other studies in the natural-language processing and computational literature of correlations between language and persuasiveness \citep
{Guerini:ProceedingsOfCicling:2008,Mitra:ProceedingsOfCscw:2014,Guerini:ProceedingsOfNaacl:2015,Tan:2016:WAI:2872427.2883081,canobasave-he:2016:N16-1}.  Hedging was one of the features examined by \citet{Tan:2016:WAI:2872427.2883081}.

\section{Conclusion}
\label{sec:conclusion}

In this paper, we took advantage of ``natural experiments'' in the same speech within meetings and proposed a computational framework for measuring the effects of subtle presentational and rhetorical (de-)emphasis.
We
applied our framework in FOMC meetings and found surprising patterns, including a positive effect of de-emphasis indicated by hedging.
Furthermore, we demonstrated how the effect of hedging is more pronounced for speakers female speakers.
This work is one step towards quantitatively understanding the effect of wording on social dynamics in meetings.
This general idea of looking at words or phrases in the same speech can spur new computational frameworks to measure the influence of language.

\section{Acknowledgments} 
We first learned of the availability of FOMC meeting transcripts from Cheryl Schonhardt-Bailey at the 2010 Text as Data meeting at Northwestern!
We thank Bitsy Perlman, Cheryl Schonhardt-Bailey, and the 2016 Text as Data attendees for helpful comments.
This work was supported by a Facebook fellowship and in part by a University of Washington Innovation Award.
\appendix

\begin{figure*}[t]
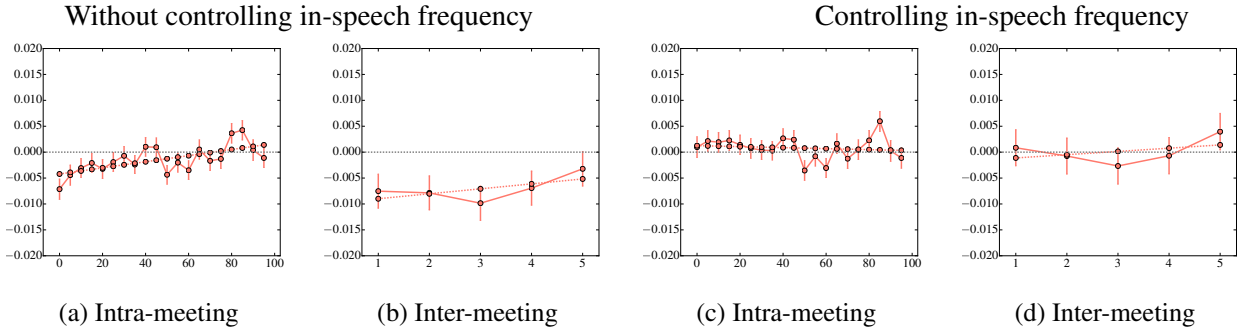

\centering
Without controlling in-speech frequency \hspace{1.4in} Controlling in-speech frequency \\
    \begin{subfigure}[t]{0.23\textwidth}
        \addFigure{\textwidth}{{content_intra_random_0.05_0.010000_noself_1006_in_out}.pdf}
        \caption{Intra-meeting}
        \label{fig:content_random}
    \end{subfigure}
    \hfill
    \begin{subfigure}[t]{0.23\textwidth}
        \addFigure{\textwidth}{{content_inter_random_0.05_0.010000_noself_1006_in_out}.pdf}
        \caption{Inter-meeting}
        \label{fig:content_inter}
    \end{subfigure}
    \hfill
    \begin{subfigure}[t]{0.23\textwidth}
        \addFigure{\textwidth}{{infreq_intra_random_0.05_0.010000_noself_1006_in_out}.pdf}
        \caption{Intra-meeting}
        \label{fig:infreq_random}
    \end{subfigure}
    \hfill
    \begin{subfigure}[t]{0.23\textwidth}
        \addFigure{\textwidth}{{infreq_inter_random_0.05_0.010000_noself_1006_in_out}.pdf}
        \caption{Inter-meeting}
        \label{fig:infreq_inter}
    \end{subfigure}
    \caption{Validation of the proposed metric using a random context ($p=0.05$).
    \label{fig:validation}}
\end{figure*}

\section{Appendix:  notes on pairing in-context vs. out-of-context words ($\MP$ from section \ref{sec:exp})}
\label{sec:validation}

Our framework considers ``natural experiments''
using word pairs drawn
 from the same speech of the same speaker.
However, there can be many intricate design choices in defining $\MP(S)$ (Equation \ref{eq:mp}) that may affect the measurement of $E(C)$.
These choices include whether to control the frequency of paired words within the speech, the part-of-speech tag of paired words, etc.
Therefore, we use a ``random'' context to validate these choices.

In order to generate a ``random'' context, we toss a coin for each word position in the speech with probability $p$ to decide whether this word position is a context cue.\footnote{We considered $p=0.05$ and $p=0.5$. The trends are similar.}
{\em Since the context is randomly selected, we expect our metric to be around 0.}
If the observed effects are different from 0, it suggests that there exists some systematic bias in the design choices.

Figure \ref{fig:validation} presents the results regarding whether to control the frequency of paired words in the same speech.
Surprisingly, given that we have already controlled for past frequency of paired words, it remains important to control for the number of times a word occurs in the speech.
Without controlling in-speech frequency, the effect is biased towards the negative side, which could have led to the wrong conclusion that a random context has negative effects on future re-occurrences of words.

We also explore other design choices that can potentially influence the metric: 1) where the speech happens in the meeting (meetings may have different stages and contexts may provide different effects in the middle of a meeting compared to in the beginning of a meeting);
2) part-of-speech tags of paired words (contexts may have different effects on words of different part-of-speech tags).
These two factors did not affect our metric.
Therefore, in the following results, we enforce that paired words have the same frequency within the same speech.

\bibliography{ref}

\begin{thebibliography}{32}
\providecommand{\natexlab}[1]{#1}
\providecommand{\url}[1]{\texttt{#1}}
\expandafter\ifx\csname urlstyle\endcsname\relax
  \providecommand{\doi}[1]{doi: #1}\else
  \providecommand{\doi}{doi: \begingroup \urlstyle{rm}\Url}\fi

\bibitem[Burgoon et~al.(1975)Burgoon, Jones, and Stewart]{burgoon1975toward}
Michael Burgoon, Stephen~B. Jones, and Diane Stewart.
\newblock {Toward a Message-centered Theory of Persuasion: Three Empirical
  Investigations of Language Intensity}.
\newblock \emph{Human Communication Research}, 1\penalty0 (3):\penalty0
  240--256, 1975.

\bibitem[Cameron(1995)]{Cameron:VerbalHygieneInRecentYearsPractices:1995}
Deborah Cameron.
\newblock The new {Pygmalion}: Verbal hygiene for women.
\newblock In \emph{Verbal Hygiene}, pages 166--211. Routledge, 1995.
\newblock URL \url{http://site.ebrary.com/id/10100241}.

\bibitem[Cano-Basave and He(2016)]{canobasave-he:2016:N16-1}
Amparo~Elizabeth Cano-Basave and Yulan He.
\newblock A study of the impact of persuasive argumentation in political
  debates.
\newblock In \emph{Proceedings of NAACL}, pages 1405--1413, June 2016.

\bibitem[Carli(2001)]{Carli:JournalOfSocialIssues:2001}
Linda~L. Carli.
\newblock Gender and social influence.
\newblock \emph{Journal of Social Issues}, 57\penalty0 (4):\penalty0 725--741,
  2001.

\bibitem[Cialdini(2009)]{Cialdini:InfluenceScienceAndPractice:2009}
Robert~B. Cialdini.
\newblock \emph{Influence: Science and Practice}.
\newblock HarperCollins, 2009.

\bibitem[Danescu-Niculescu-Mizil and
  Lee(2011)]{Danescu-Niculescu-Mizil:2011:CIC:2021096.2021105}
Cristian Danescu-Niculescu-Mizil and Lillian Lee.
\newblock Chameleons in imagined conversations: A new approach to understanding
  coordination of linguistic style in dialogs.
\newblock In \emph{Proceedings of the 2Nd Workshop on Cognitive Modeling and
  Computational Linguistics}, 2011.

\bibitem[Danescu-Niculescu-Mizil et~al.(2011)Danescu-Niculescu-Mizil, Gamon,
  and Dumais]{DanescuNiculescuMizil:ProceedingsOfWww:2011a}
Cristian Danescu-Niculescu-Mizil, Michael Gamon, and Susan Dumais.
\newblock Mark my words! {Linguistic} style accommodation in social media.
\newblock In \emph{Proceedings of WWW}, 2011.

\bibitem[Danescu-Niculescu-Mizil et~al.(2012)Danescu-Niculescu-Mizil, Lee,
  Pang, and Kleinberg]{Danescu-Niculescu-Mizil:2012:EPL:2187836.2187931}
Cristian Danescu-Niculescu-Mizil, Lillian Lee, Bo~Pang, and Jon Kleinberg.
\newblock Echoes of power: Language effects and power differences in social
  interaction.
\newblock In \emph{Proceedings of WWW}, pages 699--708, 2012.

\bibitem[Durik et~al.(2008)Durik, Britt, Reynolds, and
  Storey]{Durik:JournalOfLanguageAndSocialPsychology:2008}
Amanda~M. Durik, M.~Anne Britt, Rebecca Reynolds, and Jennifer Storey.
\newblock The effects of hedges in persuasive arguments: A nuanced analysis of
  language.
\newblock \emph{Journal of Language and Social Psychology}, 27\penalty0
  (3):\penalty0 217--234, 2008.

\bibitem[Entman(1993)]{Entman:JournalOfCommunication:1993}
Robert~M. Entman.
\newblock Framing: Toward clarification of a fractured paradigm.
\newblock \emph{Journal of Communication}, 43\penalty0 (4):\penalty0 51--58,
  1993.

\bibitem[Erickson et~al.(1978)Erickson, Lind, Johnson, and
  O'Barr]{Erickson+al:78a}
Bonnie Erickson, E.~Allan Lind, Bruce~C. Johnson, and William~M. O'Barr.
\newblock Speech style and impression formation in a court setting: The effects
  of ``powerful'' and ``powerless'' speech.
\newblock \emph{Journal of Experimental Social Psychology}, 14\penalty0
  (3):\penalty0 266 -- 279, 1978.

\bibitem[Farkas et~al.(2010)Farkas, Vincze, M\'ora, Csirik, and
  Szarvas]{Farkas+al:2010a}
Rich\'ard Farkas, Veronika Vincze, Gy\"orgy M\'ora, J\'anos Csirik, and
  Gy\"orgy Szarvas.
\newblock The {CoNLL}-2010 shared task: Learning to detect hedges and their
  scope in natural language text.
\newblock In \emph{Proceedings of the Fourteenth Conference on Computational
  Natural Language Learning---Shared Task}, pages 1--12, 2010.

\bibitem[Guerini et~al.(2008)Guerini, Strapparava, and
  Stock]{Guerini:ProceedingsOfCicling:2008}
Marco Guerini, Carlo Strapparava, and Oliviero Stock.
\newblock Trusting politicians' words (for persuasive {NLP}).
\newblock \emph{Proceedings of CICLing}, pages 263--274, 2008.

\bibitem[Guerini et~al.(2015)Guerini, Ozbal, and
  Strapparava]{Guerini:ProceedingsOfNaacl:2015}
Marco Guerini, G\"{o}zde Ozbal, and Carlo Strapparava.
\newblock Echoes of persuasion: The effect of euphony in persuasive
  communication.
\newblock In \emph{Proceedings of NAACL}, pages 1483--1493, 2015.

\bibitem[Guo et~al.(2015)Guo, Blundell, Wallach, and Heller]{guo2015bayesian}
Fangjian Guo, Charles Blundell, Hanna Wallach, and Katherine Heller.
\newblock The {Bayesian} echo chamber: Modeling social influence via linguistic
  accommodation.
\newblock In \emph{Proceedings of AISTATS}, pages 315--323, 2015.

\bibitem[Hanauer et~al.(2012)Hanauer, Liu, Mei, Manion, Balis, and
  Zheng]{hanauer2012hedging}
David~A. Hanauer, Yang Liu, Qiaozhu Mei, Frank~J. Manion, Ulysses~J. Balis, and
  Kai Zheng.
\newblock {Hedging their mets: the use of uncertainty terms in clinical
  documents and its potential implications when sharing the documents with
  patients}.
\newblock In \emph{AMIA Annual Symposium Proceedings}, 2012.

\bibitem[Hansen et~al.(2015)Hansen, McMahon, and Prat]{hansentransparency}
Stephen Hansen, Michael McMahon, and Andrea Prat.
\newblock {Transparency and deliberation within the FOMC: A computational
  linguistics approach}.
\newblock
  \url{https://www2.warwick.ac.uk/fac/soc/economics/staff/mfmcmahon/research/fomc_submission.pdf},
  2015.

\bibitem[Hyland()]{hyland1998hedging}
Ken Hyland.
\newblock \emph{Hedging in scientific research articles}.
\newblock Pragmatics and Beyond New Series.

\bibitem[Jarzabkowski and Seidl(2008)]{Jarzabkowski01112008}
Paula Jarzabkowski and David Seidl.
\newblock The role of meetings in the social practice of strategy.
\newblock \emph{Organization Studies}, 29\penalty0 (11):\penalty0 1391--1426,
  2008.

\bibitem[Meade(2005)]{meade2005fomc}
Ellen~E. Meade.
\newblock {The {FOMC}: Preferences, voting, and consensus}.
\newblock \emph{Federal Reserve Bank of St. Louis Review}, 87\penalty0
  (2):\penalty0 93--101, 2005.

\bibitem[Meade and Stasavage()]{Meade:TheEconomicJournal:2008}
Ellen~E. Meade and David Stasavage.
\newblock Publicity of debate and the incentive to dissent: Evidence from the
  {US Federal Reserve}, journal = {The Economic Journal}, year = {2008}, volume
  = {118}, number = {528}, pages = {695--717}.

\bibitem[Mitra and Gilbert(2014)]{Mitra:ProceedingsOfCscw:2014}
Tanushree Mitra and Eric Gilbert.
\newblock The language that gets people to give: Phrases that predict success
  on kickstarter.
\newblock In \emph{Proceedings of CSCW}, 2014.

\bibitem[Monroe et~al.(2008)Monroe, Colaresi, and Quinn]{Monroe21092008}
Burt~L. Monroe, Michael~P. Colaresi, and Kevin~M. Quinn.
\newblock Fightin' words: Lexical feature selection and evaluation for
  identifying the content of political conflict.
\newblock \emph{Political Analysis}, 16\penalty0 (4):\penalty0 372--403, 2008.

\bibitem[Niederhoffer and Pennebaker(2002)]{Niederhoffer+Pennebaker:2002a}
Kate~G. Niederhoffer and James~W. Pennebaker.
\newblock Linguistic style matching in social interaction.
\newblock \emph{Journal of Language and Social Psychology}, 21\penalty0
  (4):\penalty0 337--360, 2002.

\bibitem[Petty and
  Cacioppo(2012)]{Petty:CommunicationAndPersuasionCentralAndPeripheralRoutes:2012}
Richard~E. Petty and John~T. Cacioppo.
\newblock \emph{Communication and Persuasion: Central and Peripheral Routes to
  Attitude Change}.
\newblock Springer Science \& Business Media, 2012.

\bibitem[Rosa(2013)]{rosa2013financial}
Carlo Rosa.
\newblock {The financial market effect of {FOMC} minutes}.
\newblock \emph{Economic Policy Review}, 19\penalty0 (2), 2013.

\bibitem[Schonhardt-Bailey(2013)]{SchonhardtBailey:DeliberatingAmericanMonetaryPolicyATextualAnalysis:2013}
Cheryl Schonhardt-Bailey.
\newblock \emph{Deliberating American Monetary Policy: A Textual Analysis}.
\newblock MIT Press, illustrated edition, 2013.

\bibitem[Schr\"{o}der and
  Zimmer(1997)]{Schroeder:HedgingInDiscourseApproachesToTheAnalysis:1997}
Hartmut Schr\"{o}der and Dagmar Zimmer.
\newblock Hedging research in pragmatics: A bibliographical research guide to
  hedging.
\newblock In \emph{Hedging in Discourse: Approaches to the analysis of a
  pragmatic phenomenon in academic texts}, Research in Text Theory, pages
  249--271. De Gruyter, 1997.

\bibitem[Tan et~al.(2014)Tan, Lee, and Pang]{Tan:Acl:2014}
Chenhao Tan, Lillian Lee, and Bo~Pang.
\newblock The effect of wording on message propagation: Topic-and
  author-controlled natural experiments on twitter.
\newblock In \emph{Proceedings of the ACL}, pages 175--185, 2014.

\bibitem[Tan et~al.(2016)Tan, Niculae, Danescu-Niculescu-Mizil, and
  Lee]{Tan:2016:WAI:2872427.2883081}
Chenhao Tan, Vlad Niculae, Cristian Danescu-Niculescu-Mizil, and Lillian Lee.
\newblock Winning arguments: Interaction dynamics and persuasion strategies in
  good-faith online discussions.
\newblock In \emph{Proceedings of WWW}, pages 613--624, 2016.

\bibitem[Zhang et~al.(2016)Zhang, Kumar, Ravi, and
  Danescu-Niculescu-Mizil]{Zhang:ProceedingsOfNaacl:2016}
Justine Zhang, Ravi Kumar, Sujith Ravi, and Cristian Danescu-Niculescu-Mizil.
\newblock {Conversational flow in Oxford-style debates}.
\newblock In \emph{Proceedings of NAACL (short papers)}, 2016.

\bibitem[Zirn et~al.(2015)Zirn, Meusel, and Stuckenschmidt]{zirnlost}
Cäcilia Zirn, Robert Meusel, and Heiner Stuckenschmidt.
\newblock Lost in discussion? {Tracking} opinion groups in complex political
  discussions by the example of the fomc meeting transcriptions.
\newblock In \emph{Proceedings of RANLP}, pages 747--753, 2015.

\end{thebibliography}
\bibliographystyle{plainnat} %

\end{document}